\documentclass[12pt]{article}
\usepackage{graphicx}
\usepackage{amssymb}
\usepackage{amsbsy}
\usepackage{bm}
\usepackage{graphicx}   
\usepackage{psfrag}
\usepackage{dcolumn}
\usepackage{amssymb}
\usepackage{bm}
\usepackage{subfigure}
\usepackage{caption}
\usepackage{amsmath}
\usepackage{amstext}
\usepackage{amsthm}
\usepackage{amssymb}
\usepackage{amsopn}
\usepackage{mathrsfs} 
\usepackage{dsfont} 
\usepackage[bottom]{footmisc}
\usepackage{indentfirst}
\usepackage{upref}
\usepackage{cite}
\usepackage{units}
\usepackage{fancybox}
\usepackage
{floatflt}
\numberwithin{equation}{section}
\usepackage{ifpdf}
\usepackage{caption}
\usepackage{subfigure}
\usepackage{amsmath}
\usepackage{amssymb}
\usepackage{cite}
\usepackage{psfrag}
\usepackage{fontenc}
\usepackage{times}
\usepackage{mathptmx}
\usepackage{graphicx}
\newcommand{\Lr}{{\rm L}}
\newcommand{\Dr}{{\rm D}}
\newcommand{\Br}{{\rm B}}
\newcommand{\Gr}{{\rm G}}
\newcommand{\dr}{{\rm d}}

\long\def\symbolfootnote[#1]#2{\begingroup%
\def\thefootnote{\fnsymbol{footnote}}\footnote[#1]{#2}\endgroup}

\begin{document}
\begin{titlepage} 
\renewcommand{\baselinestretch}{1.1}
\small\normalsize
\begin{flushright}
MZ-TH/11-40
\end{flushright}

\vspace{0.1cm}
\vspace{0.1cm}
\vspace{0.1cm}
\vspace{0.1cm}
\vspace{1cm}\vspace{1cm}
\begin{center}   

{\Large \textbf{Running Immirzi Parameter \\and Asymptotic Safety} \renewcommand{\thefootnote}{\fnsymbol{footnote}}\footnote[1]{Talk given by Martin Reuter at Corfu Summer Institute on Elementary Particles and Physics - Workshop on Non Commutative Field Theory and Gravity, September 8-12, 2010}\renewcommand{\thefootnote}{\arabic{footnote}}}

\vspace{1.4cm}
{\large J.-E.~Daum and M.~Reuter}\\

\vspace{0.7cm}
\noindent
\textit{Institute of Physics, University of Mainz\\
Staudingerweg 7, D--55099 Mainz, Germany}\\

\end{center}

\vspace*{0.6cm}

\begin{abstract}
\vspace{12pt}
We explore the renormalization group (RG) properties of quantum gravity, using the vielbein and the spin connection as the fundamental field variables. We require the effective action to be invariant under the semidirect product of spacetime diffeomorphisms and local frame rotations. Starting from the corresponding functional integral we review the construction of an appropriate theory space and an exact funtional RG equation operating on it. We then solve this equation on a truncated space defined by a three parameter family of Holst-type actions which involve a running Immirzi parameter. We find evidence for the existence of an asymptotically safe fundamental theory. It is probably inequivalent to metric quantum gravity constructed in the same way.
\end{abstract}

\end{titlepage}

\section{Introduction}\label{s1}
During the past decade the gravitational effective average action \cite{mr} has been used in a number of studies trying to understand the renormalization behavior of Quantum Einstein Gravity (QEG) at a nonperturbative level. An important motivation was Weinberg's idea of Asymptotic Safety \cite{wein} according to which gravity might be nonperturbatively renormalizable and predictive if there exists a nontrivial renormalization group (RG) fixed point with a finite dimensional ultraviolet critical manifold at which the infinite cutoff limit can be taken. All investigations carried out so far point in the direction that the RG flow of the effective average action does indeed possess an RG fixed point with the desired properties \cite{QEG,prop}. 

In a nutshell, the Asymptotic Safety program can be summarized roughly as follows \cite{reviews}: 

\noindent {\bf (i)} Fix a set of fields $\Phi$ carrying the gravitational degrees of freedom. 

\noindent {\bf (ii)} Pick a group ${\bf G}$ of gauge or symmetry transformations acting on $\Phi$. 

\noindent {\bf (iii)} Define a ``theory space'' consisting of all action functionals invariant under ${\bf G}$, i.\,e. ${\cal T} \equiv \{A[\Phi] | A ~ \mbox{invariant under} ~ {\bf G}\}$. 

\noindent {\bf (iv)} Fix a coarse graining scheme on ${\cal T}$, a background covariant continuum analogue of the Kadanoff-Wilson block spin idea \cite{mr}. 

\noindent {\bf (v)} Compute the corresponding ``RG flow'' $({\cal T}, \,\beta)$ where $\beta$ is the vector field on ${\cal T}$ obtained by applying an infinitesimal coarse graining step $A \mapsto A + \beta (A)$ to all actions, and interpret $\beta (A)$ as an element of the tangent space ${\rm T}_A {\cal T}$. 

\noindent {\bf (vi)} Compute the resulting ``RG trajectories'' $\Gamma_\bullet : \mathbb{R} \rightarrow {\cal T}, \: k \mapsto \Gamma_k$ as the integral curves of $\beta$, i.\,e. solve the ``flow equation'' or ``functional RG equation'' (FRGE) $\frac{{\rm d}}{{\rm d}\,{\rm ln}\,k} \Gamma_k = \beta (\Gamma_k)$. For the gravitational effective average action the coarse graining scheme is concretely defined by setting $\beta(\Gamma_k) = \frac{1}{2} {\rm STr} \big[(\Gamma_k^{(2)} + {\cal R}_k)^{-1} k \partial_k {\cal R}_k\big]$ where $\Gamma_k^{(2)}$ is the functional Hessian of $\Gamma_k$ and ${\cal R}_k$ a cutoff kernel \cite{mr}. For this choice, $\Gamma_{k \to 0}$ coincides with the ordinary effective action, and $\Gamma_{k \to \infty}$ is closely related to the bare action $S$ \cite{elisa}. 

\noindent {\bf (vii)} Determine the fixed points of the flow, i.\,e. try to solve $\beta (A_\star) = 0$. 

\noindent {\bf (viii)} If there exists a fixed point, linearize the flow about $A_\star$ and solve the linear system $\frac{{\rm d}}{{\rm d}\,{\rm ln}\,k} \delta \Gamma_k = {\cal B}\:\delta\Gamma_k$ where ${\cal B}$ is the Jacobi matrix of $\beta$ at $A_\star$. Its (negative) eigenvalues are the ``critical exponents'' $\Theta_\alpha$ and its eigenvectors are the ``scaling fields'' behaving as $(\delta\Gamma_k)_\alpha \propto k^{-\Theta_\alpha}$ near $A_\star$. We say a scaling field is relevant (irrelevant) if it increases (decreases) when the mass scale $k$ is lowered. 

\noindent {\bf (ix)} Try to find complete RG trajectories, i.\,e. trajectories for which both the ultraviolet (UV) limit $k \to \infty$ and the infrared (IR) limit $k \to 0$ exist. Every such trajectory defines a quantum field theory (in the sense of {\em all} modes of the fundamental field being integrated out). The Asymptotic Safety idea consists in taking the UV limit at a non-Gaussian fixed point (NGFP), i.\,e. to ensure the UV-regularity of the trajectory by arranging it to hit a fixed point asymtotically: $\Gamma_k \to A_\star$ for $k \to \infty$.

For the case where the fundamental field is assumed to be the spacetime metric $g_{\mu\nu}$\footnote{Together with a background metric and Faddeev-Popov ghosts, for technichal reasons.} the viability of the above program has been tested to some extent and significant evidence for the existence of an appropriate NGFP was found. However, it is clear that other choices are equally plausible here. In Einstein-Cartan gravity, for example, the field variables are constituted by the vielbein $e^a_{~\mu}$ and the spin connection $\omega^{ab}_{~~\mu}$ assuming values in the Lie algebra of the Lorentz group. Since $\omega^{ab}_{~~\mu}$ can carry spacetime torsion, Einstein-Cartan gravity has in general more degrees of freedom than metric based general relativity; only in absence of ``spinning'' matter the theories happen to possess equivalent classical field equations. The dynamics of Einstein-Cartan theory is encoded in the Hilbert-Palatini action $S_{\rm HP} [e, \omega]$ which is of first order in the spacetime derivatives. 

Interest in the field variables $(e^a_{~\mu},\, \omega^{ab}_{~~\mu})$ stems also from several modern developments towards the quantization of gravity which use related variables. This includes canonical quantum gravity with Ashtekar's variables \cite{A,R}, loop quantum gravity \cite{T}, spin foam models \cite{SF}, and group field theory \cite{GFT}. Here the Hilbert-Palatini action is usually generalized to the so-called Holst action $S_{\rm Ho}$ \cite{soeren} which contains an additional term, specific to four dimensions,  whose associated coupling is the Immirzi parameter $\gamma$. While the classical vacuum field equations implied by $S_{\rm Ho}$ are independent of the dimensionless number $\gamma$, the corresponding quantum theory seems to depend on it. Within Loop Quantum Gravity (LQG), $\gamma$ enters the eigenvalues of area and volume operators, as well as the formula for the entropy of black holes \cite{R}.

In order to explore the possibility of constructing asymptotically safe quantum theories of gravity in which $e^a_{~\mu}$ and $\omega^{ab}_{~~\mu}$ serve as the fundamental field variables, we perform a first analysis of the Wilsonian RG flow on the corresponding theory space. In Section \ref{s2} of this contribution we describe the flow equation used and the theory space it acts on, and in Section \ref{s3} we present the results obtained. Finally, Section \ref{s4} provides a short conclusion. For further details, we refer to \cite{eomega2}.         

%
%
%
\section{Construction of Theory Space and Flow Equation}\label{s2}

\subsection{Fields and gauge invariances}\label{ss21}

We start out from an a priori formal functional integral ${\cal Z} = \int {\cal D}\hat{e}^a_{~\mu}\:{\cal D}\hat{\omega}^{ab}_{~~\mu}\:{\rm exp}\big\{-S[\hat{e}, \hat{\omega}]\big\}$, where the quantum fields $\hat{e}^a_{~\mu}$ and $\hat{\omega}^{ab}_{~~\mu}$ are defined on a fixed (differentiable) manifold without boundary, ${\cal M}$, and the bare action $S$ is invariant both under diffeomorphisms ${\sf Diff} ({\cal M})$ and local Lorentz rotations. We consider the euclidean form of the theory, so that the relevant group of gauge transformations is the semidirect product ${\bf G} = {\sf Diff}({\cal M}) \ltimes {\sf O}(4)_{\rm loc}$. For every given co-frame $\hat{e}^a_{~\mu}$ and ${\sf o}(4)$-valued connection $\hat{\omega}^{ab}_{~~\mu}$ on ${\cal M}$ we are provided with an ${\sf O}(4)$-covariant derivative 
\begin{align}
\hat{\nabla}_\mu \equiv \partial_\mu + \frac{1}{2} \hat{\omega}^{ab}_{~~\mu} M_{ab}
\end{align}
where $M_{ab}$ are the generators in the corresponding representation, and with the associated curvature and torsion tensors 
\begin{align}
\hat{F}^{ab}_{~~\mu\nu} \equiv \partial_\mu \hat{\omega}^{ab}_{~~\nu} + \hat{\omega}^a_{~c\mu}\hat{\omega}^{cb}_{~~\nu} - (\mu \leftrightarrow \nu)
\end{align} 
and 
\begin{align}
\hat{T}^a_{\mu\nu} \equiv \partial_\mu \hat{e}^a_{~\nu} + \hat{\omega}^a_{~c\mu}\hat{e}^c_{~\nu} - (\mu \leftrightarrow \nu)\,, 
\end{align}
respectively. 

Under ${\sf O}(4)_{\rm loc}$ we have 
\begin{equation}\label{o4-undecomp}
\mbox{$\begin{array}{lcl}\displaystyle \delta_{\rm L}\ (\lambda) \hat{e}^a_{~\mu} &=& \lambda^a_{~b} \hat{e}^b_{~\mu}\:,\\
\displaystyle \delta_\Lr (\lambda) \hat{\omega}^{ab}_{~~\mu} &=& - \partial_\mu \lambda^{ab} + \lambda^a_{~c}\hat{\omega}^{cb}_{~~\mu} + \lambda^b_{~c}\hat{\omega}^{ac}_{~~\mu} \equiv - \hat{\nabla}_\mu \lambda^{ab} \end{array}$}
\end{equation}
where $\hat{\nabla}$ is the ${\sf O}(4)$ covariant derivative pertaining to $\hat{\omega}^{ab}_{~~\mu}$, while under diffeomorphisms 
\begin{equation}\label{diffeomorphisms-undecomp}
\mbox{$\begin{array}{lcl}\displaystyle \delta_\Dr (w) \hat{e}^a_{~\mu} &=& {\cal L}_w \hat{e}^a_{~\mu}\:, \\
\displaystyle \delta_\Dr (w) \hat{\omega}^{ab}_{~~\mu} &=& {\cal L}_w \hat{\omega}^{ab}_{~\mu} \end{array}$}
\end{equation}
where ${\cal L}_w$ denotes the Lie derivative along the generating vector field $w$. 

In the quantum theory we also need to consider diffeomorphism ghosts and antighosts, ${\cal C}^\mu$ and $\bar{{\cal C}}_\mu$, respectively, and likewise $\Sigma^{ab}$ and $\bar{\Sigma}_{ab}$ for the local ${\sf O}(4)$ transformations. We require all ghost and antighost fields to transform under ${\sf Diff} ({\cal M})$ and ${\sf O (4)}_{\rm loc}$ as tensors of the corresponding type.

It then follows that the algebra of all gauge transformations is given by
\begin{equation}\label{field-algebra}
\fbox{$ \begin{array}{lcl}\displaystyle [\delta_\Dr (w_1), \delta_\Dr (w_2)] \Phi &=& \delta_\Dr ([w_1, w_2]) \Phi \\
\displaystyle [\delta_\Lr (\lambda_1), \delta_\Lr (\lambda_2)] \Phi &=& \delta_\Lr ([\lambda_1, \lambda_2]) \Phi \\
\displaystyle [\delta_\Dr (w), \delta_\Lr (\lambda)] \Phi &=& \delta_\Lr ({\cal L}_w \lambda) \Phi \vspace{0.25cm}\\
&&\hspace{-3.35cm}\displaystyle \forall\:\Phi \in \{\hat{e}^a_{~\mu}, \hat{\omega}^{ab}_{~~\mu}, {\cal C}^\mu, \bar{{\cal C}}_\mu, \Sigma^{ab}, \Sigma_{ab}\} \end{array} $} \end{equation}
Here $[w_1, w_2]$ denotes the Lie bracket of the vector fields $w_1$ and $w_2$, and $[\lambda_1, \lambda_2]$ is the commutator of two matrices. The algebra is a semidirect product ${\sf Diff} ({\cal M}) \ltimes {\sf O(4)}_{\rm loc}$ with the local Lorentz transformations playing the role of the invariant subalgebra.

In order to implement the gauge transformations on the space of functionals $A [\hat{e}^a_{~\mu}, \hat{\omega}^{ab}_{~~\mu}, {\cal C}^\mu, \bar{{\cal C}}_\mu,\\ \Sigma^{ab}, \bar{\Sigma}_{ab}]$ we introduce the corresponding Ward operators ${\cal W}_{\Dr}$ and ${\cal W}_{\Lr}$ such that $\delta_{\Dr,\,\Lr}\,A = - {\cal W}_{\Dr,\,\Lr}\,A$ to linear order in the transformation parameters. Explicitly, 
\begin{align}\label{W-D}
& {\cal W}_{\Dr} (w) = - \int \dr^4 x \Bigl(\delta_\Dr (w) \hat{e}^a_{~\mu} (x) \frac{\delta}{\delta \hat{e}^a_{~\mu} (x)} + \delta_\Dr (w) \hat{\omega}^{ab}_{~~\mu} (x) \frac{\delta}{\delta \hat{\omega}^{ab}_{~~\mu} (x)}\nonumber\\ 
& \hspace{1.5cm} + \delta_\Dr (w) {\cal C}^\mu (x) \frac{\delta}{\delta {\cal C}^\mu (x)} + \delta_\Dr (w) \bar{{\cal C}}_\mu (x) \frac{\delta}{\delta {\bar{\cal C}}_\mu (x)}\nonumber\\
& \hspace{1.5cm} + \delta_\Dr (w) \Sigma^{ab} (x) \frac{\delta}{\delta \Sigma^{ab} (x)} + \delta_\Dr (w) \bar{\Sigma}_{ab} (x) \frac{\delta}{\delta \bar{\Sigma}_{ab} (x)}\Bigr)
\end{align}
and analogously for ${\cal W}_{\Lr}$.  The Ward operators satisfy
\begin{equation}\label{ward-algebra}
\fbox{$ \begin{array}{lcl}\displaystyle [{\cal W}_{\Dr} (w_1), {\cal W}_{\Dr} (w_2)] &=& {\cal W}_{\Dr} ([w_1, w_2]) \\
\displaystyle [{\cal W}_{\Lr} (\lambda_1), {\cal W}_{\Lr} (\lambda_2)] &=& {\cal W}_{\Lr} ([\lambda_1, \lambda_2]) \\
\displaystyle [{\cal W}_\Dr (w), {\cal W}_\Lr (\lambda)] &=& {\cal W}_\Lr ({\cal L}_w \lambda) \end{array}$}
\end{equation}
Gauge invariant functionals $A[\hat{e}, \hat{\omega}, {\cal C}, \bar{{\cal C}}, \Sigma, \bar{\Sigma}]$ are characterized by the conditions ${\cal W}_{~\Dr} (w) \, A = 0 = {\cal W}_{~\Lr} (\lambda) \, A$ for all $w$ and $\lambda$. 

In order to ultimatively arrive at a functional integral and a flow equation with the desired invariance properties it is important to notice that the (ordinary) diffeomorphisms $\delta_\Dr (w)$ are not covariant under ${\sf O(4)}_{\rm loc}$. This is obvious from the fact that the Lie derivative involves partial rather than ${\sf O(4)}$-covariant derivatives. We can, however, covariantize the diffeomorphisms by combining them with an appropriate ${\sf O(4)}$ transformation. Introducing
\begin{align}
\widetilde{\delta_\Dr} (w) \equiv \delta_\Dr (w) + \delta_\Lr (w \cdot \hat{\omega})
\end{align}
where $(w \cdot \omega)^{ab} \equiv w^\mu \omega^{ab}_{~~\mu}$, the action of the modified diffeomorphisms $\widetilde{\delta_\Dr}$ involves covariant derivatives $\hat{\nabla}_\mu$ in place of $\partial_\mu$:
\begin{equation}
\mbox{$\begin{array}{lcl}\displaystyle \widetilde{\delta_\Dr} (w) \hat{e}^a_{~\mu} &=& w^\rho \hat{\nabla}_\rho \hat{e}^a_{~\mu} + (\hat{\nabla}_\mu w^\rho) \hat{e}^a_{~\rho}\\
\displaystyle \widetilde{\delta_\Dr} (w) \hat{\omega}^{ab}_{~~\mu} &=& - \hat{F}^{ab}_{~~\mu\nu} w^\rho\\
\displaystyle \widetilde{\delta_\Dr} (w) {\cal C}^\mu &=& w^\rho \hat{\nabla}_\rho {\cal C}^\mu - (\hat{\nabla}_\rho w^\mu) {\cal C}^\rho = w^\rho \partial_\rho {\cal C}^\mu - (\partial_\rho w^\mu) {\cal C}^\rho \\
\displaystyle \widetilde{\delta_\Dr} (w) \bar{{\cal C}}_\mu &=& w^\rho \hat{\nabla}_\rho \bar{{\cal C}}_\mu + (\hat{\nabla}_\mu w^\rho) \bar{{\cal C}}_\rho\\  
\displaystyle \widetilde{\delta_\Dr} (w) \Sigma^{ab} &=& w^\rho \hat{\nabla}_\rho \Sigma^{ab}\\
\displaystyle \widetilde{\delta_\Dr} (w) \bar{\Sigma}_{ab} &=& w^\rho \hat{\nabla}_\rho \bar{\Sigma}_{ab} \end{array}$}
\end{equation}
Associating in the usual way Ward operators $\widetilde{{\cal W}_\Dr} (w)$ to the modified diffeomorphisms leads to the following covariantized form of the gauge algebra:
\begin{equation}\label{ward-algebra-new}
\fbox{$ \begin{array}{lcl}\displaystyle [\widetilde{{\cal W}_\Dr} (w_1), \widetilde{{\cal W}_\Dr} (w_2)] &=& \widetilde{{\cal W}_\Dr} ([w_1, w_2]) - {\cal W}_\Lr (w_1 w_2 \cdot \hat{F}) \\
\displaystyle [{\cal W}_\Lr (\lambda_1), {\cal W}_\Lr (\lambda_2)] &=& {\cal W}_\Lr ([\lambda_1, \lambda_2]) \\
\displaystyle [\widetilde{{\cal W}_\Dr} (w), {\cal W}_\Lr (\lambda)] &=& 0 \end{array}$}
\end{equation}
Here $(w_1 w_2 \cdot \hat{F})^{ab} \equiv w_1^\mu w_2^\nu \, \hat{F}^{ab}_{~~\mu\nu}$. Note that while the modified diffeomorphisms commute with local Lorentz transformations, they no longer close among themselves; their commutator contains an ${\sf O (4)}_{\rm loc}$ transformation whose parameter involves $\hat{F}$, the curvature of $\hat{\omega}$.

Note also that gauge invariant functionals $A$ are equivalently characterized by the conditions $\widetilde{{\cal W}_\Dr} (w) \, A = 0 = {\cal W}_\Lr (\lambda) \, A$ for all $w$ and $\lambda$.

\subsection{Gauge fixing and modified diffeomorphisms}\label{ss22}

In order to arrive at a functional integral which can be computed (actually {\em defined}) by means of a functional RG flow we introduce arbitrary background fields\footnote{The background vielbein $\bar{e}^a_{~\mu}$ is assumed to be nondegenerate. As a result, it gives rise to a welldefined inverse $(\bar{e}_a^{~\mu}) \equiv (\bar{e}^a_{~\mu})^{-1}$, to a nondegenerate background metric $\bar{g}_{\mu\nu} \equiv \bar{e}^a_{~\mu} \bar{e}^b_{~\nu} \delta_{ab}$, and to a completely covariant derivative $\bar{D} \equiv \partial + \bar{\omega} + \bar{\Gamma} \equiv \bar{\nabla} + \bar{\Gamma}$ where $\bar{\Gamma} \equiv \bar{\Gamma} (\bar{e}, \bar{\omega})$ is fixed by the requirement $\bar{D}_\mu \bar{e}^a_{~\nu} = 0$. Coordinate (frame) indices are denoted by greek (latin) letters. While coordinate indices are lowered and raised by means of $\bar{g}_{\mu\nu}$ and its inverse $\bar{g}^{\mu\nu}$, we lower and raise frame indices with $\delta_{ab}$ and $\delta^{ab}$, respectively.} $\bar{e}^a_{~\mu}$ and $\bar{\omega}^{ab}_{~~\mu}$, decompose the variables of integration as $\hat{e}^a_{~\mu} \equiv \bar{e}^a_{~\mu} + \varepsilon^a_{~\mu}$, $\hat{\omega}^{ab}_{~~\mu} \equiv \bar{\omega}^{ab}_{~~\mu} + \tau^{ab}_{~~\mu}$, and perform a background covariant gauge fixing. This leads to a functional integral of the form 
\begin{eqnarray}
{\cal Z} &=& \int{\cal D}\varepsilon^a_{~\mu}\:{\cal D}\tau^{ab}_{~~\mu}\:{\rm exp}\big\{- S[\bar{e} + \varepsilon, \bar{\omega} + \tau] - S_{\rm gf} [\varepsilon, \tau; \bar{e}, \bar{\omega}]\big\}\nonumber\\
&& \times \int{\cal D}{\cal C}^\mu\:{\cal D}\bar{{\cal C}}_\mu\:{\cal D}\Sigma^{ab}\:{\cal D}\bar{\Sigma}_{ab} \:{\rm exp}\big\{- S_{\rm gh}\big\} \label{z-functional}
\end{eqnarray}
Here $S_{\rm gf}$ and $S_{\rm gh}$ denote the gauge fixing and corresponding ghost action, respectively, ${\cal C}^\mu$ and $\bar{{\cal C}}_\mu$ are the diffeomorphism ghosts, and similarly $\Sigma^{ab}$ and $\bar{\Sigma}_{ab}$ are those related to the local ${\sf O}(4)$. With $G$ denoting Newton's constant, the gauge fixing is of the form
\begin{equation}\label{gf}
\begin{aligned}
S_{\rm gf} = & \frac{1}{2 \alpha_{\rm D}\cdot 16 \pi G}\int {\rm d}^4 x\:\bar{e}\:\bar{g}^{\mu\nu}\:{\cal F}_\mu {\cal F}_\nu \\
& + \frac{1}{2 \alpha_{\rm L}} \int {\rm d}^4 x\:\bar{e}\:{\cal G}^{ab} {\cal G}_{ab}
\end{aligned}
\end{equation}
where ${\cal F}_\mu$ and ${\cal G}^{ab}$ break the ${\sf Diff} ({\cal M})$ and ${\sf O}(4)_{\rm loc}$ gauge invariance, respectively. However, in order to ultimately arrive at a ${\sf Diff}({\cal M}) \ltimes {\sf O}(4)_{\rm loc}$ invariant effective average action we employ gauge conditions of the ``background type'' so that $S_{\rm gf} [\varepsilon, \tau; \bar{e}, \bar{\omega}]$ is invariant under the combined background gauge transformations $\delta^{\rm B}_{{\rm D,\,L}}$ acting on both $(\varepsilon, \,\tau)$ and $(\bar{e}, \,\bar{\omega})$ while, of course, it is not invariant under the ``true'' (or ``quantum'') gauge transformations, denoted by $\delta^{\rm G}_{\:{\rm D}}$ and $\delta^{\rm G}_{\:\,{\rm L}}$, respectively. 

The true diffeomorphisms read
\begin{equation}
\mbox{$\begin{array}{lcl}\displaystyle \delta^\Gr_{\:\Dr} (w) \bar{e}^a_{~\mu} &=& 0\:,\\
\displaystyle \delta^\Gr_{\:\Dr} (w) \varepsilon^a_{~\mu} &=& {\cal L}_w (\bar{e}^a_{~\mu} + \varepsilon^a_{~\mu})\:,\\
\displaystyle \delta^\Gr_{\:\Dr} (w) \bar{\omega}^{ab}_{~~\mu}&=& 0\:,\\
\displaystyle \delta^\Gr_{\:\Dr} (w) \tau^{ab}_{~~\mu} &=& {\cal L}_w (\bar{\omega}^{ab}_{~~\mu} + \tau^{ab}_{~~\mu}) \end{array}$}
\end{equation}
and their ${\sf O(4)}$ counterparts are     
\begin{equation}
\mbox{$\begin{array}{lcl}\displaystyle \delta^\Gr_{\:\,\Lr} (\lambda) \bar{e}^a_{~\mu} &=& 0\:,\\
\displaystyle \delta^\Gr_{\:\,\Lr} (\lambda) \varepsilon^a_{~\mu} &=& \lambda^a_{~b} (\bar{e}^b_{~\mu} + \varepsilon^b_{~\mu})\:,\\
\displaystyle \delta^\Gr_{\:\,\Lr} (\lambda) \bar{\omega}^{ab}_{~~\mu} &=& 0\:,\\
\displaystyle \delta^\Gr_{\:\,\Lr} (\lambda) \tau^{ab}_{~~\mu} &=& - \partial_\mu \lambda^{ab} + \lambda^a_{~c} (\bar{\omega}^{cb}_{~~\mu} + \tau^{cb}_{~~\mu}) + \lambda^b_{~c} (\bar{\omega}^{ac}_{~~\mu} + \tau^{ac}_{~~\mu}) \:. \end{array}$}
\end{equation}
On the other hand, the background diffeomorphisms act as
\begin{equation}
\mbox{$\begin{array}{lcl}\displaystyle \delta^\Br_{\:\Dr} (w) \bar{e}^a_{~\mu} &=& {\cal L}_w \bar{e}^a_{~\mu}\:,\\
\displaystyle \delta^\Br_{\:\Dr} (w) \varepsilon^a_{~\mu} &=& {\cal L}_w \varepsilon^a_{~\mu}\:,\\
\displaystyle \delta^\Br_{\:\Dr} (w) \bar{\omega}^{ab}_{~~\mu} &=& {\cal L}_w \bar{\omega}^{ab}_{~~\mu} \:,\\
\displaystyle \delta^\Br_{\:\Dr} (w) \tau^{ab}_{~~\mu} &=& {\cal L}_w  \tau^{ab}_{~~\mu} \end{array}$}
\end{equation}
and the background ${\sf O(4)}$ transformations are
\begin{equation}
\mbox{$\begin{array}{lcl}\displaystyle \delta^\Br_{\:\,\Lr} (\lambda) \bar{e}^a_{~\mu} &=& \lambda^a_{~b} \bar{e}^b_{~\mu} \:,\\
\displaystyle \delta^\Br_{\:\,\Lr} (\lambda) \varepsilon^a_{~\mu} &=& \lambda^a_{~b} \varepsilon^b_{~\mu}\:,\\
\displaystyle \delta^\Br_{\:\,\Lr} (\lambda) \bar{\omega}^{ab}_{~~\mu} &=& - \partial_\mu \lambda^{ab} + \lambda^a_{~c} \bar{\omega}^{cb}_{~~\mu} + \lambda^b_{~c} \bar{\omega}^{ac}_{~~\mu} \equiv - \bar{\nabla}_\mu \lambda^{ab}\:,\\
\displaystyle \delta^\Br_{\:\,\Lr} (\lambda) \tau^{ab}_{~~\mu} &=& \lambda^a_{~c} \tau^{cb}_{~~\mu} + \lambda^b_{~c} \tau^{ac}_{~~\mu} \end{array}$} 
\end{equation}
where $\bar{\nabla}$ denotes the ${\sf O(4)}$ covariant derivative constructed from $\bar{\omega}^{ab}_{~~\mu}$.

Since no background split is introduced for the ghost fields, their true and background gauge transformations happen to coincide. We require a tensorial transformation law corresponding to their index structure:
\begin{equation}
\mbox{$\begin{array}{lcl}\displaystyle \delta^\Br_{\:\Dr} (w) {\cal C}^\mu &=& \delta^\Gr_{\:\Dr} (w) {\cal C}^\mu = {\cal L}_w {\cal C}^\mu \:,\hspace{0.27cm} \delta^\Br_{\:\,\Lr} (\lambda) {\cal C}^\mu \, = \delta^\Gr_{\:\,\Lr} (\lambda) {\cal C}^\mu = 0 \:,\\
\displaystyle \delta^\Br_{\:\Dr} (w) \bar{{\cal C}}_\mu &=& \delta^\Gr_{\:\Dr} (w) \bar{{\cal C}}_\mu = {\cal L}_w \bar{{\cal C}}_\mu \:,\hspace{0.4cm} \delta^\Br_{\:\,\Lr} (\lambda) \bar{{\cal C}}_\mu \,\, = \delta^\Gr_{\:\,\Lr} (\lambda) \bar{{\cal C}}_\mu = 0 \:,\\
\displaystyle \delta^\Br_{\:\Dr} (w) \Sigma^{ab} &=& \delta^\Gr_{\:\Dr} (w) \Sigma^{ab} = {\cal L}_w \Sigma^{ab} \:,\hspace{0.3cm} \delta^\Br_{\:\,\Lr} (\lambda) \Sigma^{ab} = \delta^\Gr_{\:\,\Lr} (\lambda) \Sigma^{ab} = {\lambda^a}_c \Sigma^{cb} + {\lambda^b}_c \Sigma^{ac} \:,\\
\displaystyle \delta^\Br_{\:\Dr} (w) \bar{\Sigma}_{ab} &=& \delta^\Gr_{\:\Dr} (w) \bar{\Sigma}_{ab} = {\cal L}_w \bar{\Sigma}_{ab} \:,\hspace{0.3cm} \delta^\Br_{\:\,\Lr} (\lambda) \bar{\Sigma}_{ab} = \delta^\Gr_{\:\,\Lr} (\lambda) \bar{\Sigma}_{ab} = {\lambda_a}^c \bar{\Sigma}_{cb} + {\lambda_b}^c \bar{\Sigma}_{ac} \:.\end{array}$}
\end{equation}

Introducing Ward operators ${\cal W}^\Br_{~\Dr}$, ${\cal W}^\Br_{~\Lr}$ for the background gauge transformations, and ${\cal W}^\Gr_{~\Dr}$, ${\cal W}^\Gr_{~\Lr}$ for the ``gauge'' or ``true'' ones we can verify that the former satisfy the algebra
\begin{equation}\label{ward-B-algebra}
\fbox{$ \begin{array}{lcl}\displaystyle [{\cal W}^\Br_{~\Dr} (w_1), {\cal W}^\Br_{~\Dr} (w_2)] &=& {\cal W}^\Br_{~\Dr} ([w_1, w_2]) \\
\displaystyle [{\cal W}^\Br_{~\Lr} (\lambda_1), {\cal W}^\Br_{~\Lr} (\lambda_2)] &=& {\cal W}^\Br_{~\Lr} ([\lambda_1, \lambda_2]) \\
\displaystyle [{\cal W}^\Br_{~\Dr} (w), {\cal W}^\Br_{~\Lr} (\lambda)] &=& {\cal W}^\Br_{~\Lr} ({\cal L}_w \lambda) \end{array}$}
\end{equation}
while the latter obey the relations
\begin{equation}\label{ward-G-algebra}
\fbox{$ \begin{array}{lcl}\displaystyle [{\cal W}^\Gr_{~\Dr} (w_1), {\cal W}^\Gr_{~\Dr} (w_2)] &=& {\cal W}^\Gr_{~\Dr} ([w_1, w_2]) \\
\displaystyle [{\cal W}^\Gr_{~\Lr} (\lambda_1), {\cal W}^\Gr_{~\Lr} (\lambda_2)] &=& {\cal W}^\Gr_{~\Lr} ([\lambda_1, \lambda_2]) \\
\displaystyle [{\cal W}^\Gr_{~\Dr} (w), {\cal W}^\Gr_{~\Lr} (\lambda)] &=& {\cal W}^\Gr_{~\Lr} ({\cal L}_w \lambda) \end{array}$}
\end{equation}
Like their precursors before the background split, these commutation relations are not ${\sf O(4)}_{\rm loc}$ covariant. 

Within the background field setting we define modified diffeomorphisms according to
\begin{align}
\widetilde{\widetilde{\delta^\Br_{\:\Dr}}} (w) \equiv \delta^\Br_{\:\Dr} (w) + \delta^\Br_{\:\,\Lr} (w\cdot \bar{\omega})\:, \\
\widetilde{\widetilde{\delta^\Gr_{\:\Dr}}} (w) \equiv \delta^\Gr_{\:\Dr} (w) + \delta^\Gr_{\:\,\Lr} (w\cdot \bar{\omega})\:. 
\end{align}
In terms of their Ward operators, the modified background diffeomorphisms satisfy the commutation relations
\begin{equation}\label{ward-algebra-back-new}
\fbox{$ \begin{array}{lcl}\displaystyle [\widetilde{\widetilde{{\cal W}^\Br_{~\Dr}}} (w_1), \widetilde{\widetilde{{\cal W}^\Br_{~\Dr}}} (w_2)] &=& \widetilde{\widetilde{{\cal W}^\Br_{~\Dr}}} ([w_1, w_2]) - {\cal W}^\Br_{~\Lr} (w_1 w_2 \cdot \bar{F}) \\
\displaystyle [{\cal W}^\Br_{~\Lr} (\lambda_1), {\cal W}^\Br_{~\Lr} (\lambda_2)] &=& {\cal W}^\Br_{~\Lr} ([\lambda_1, \lambda_2]) \\
\displaystyle [\widetilde{\widetilde{{\cal W}^\Br_{~\Dr}}} (w), {\cal W}^\Br_{~\Lr} (\lambda)] &=& 0 \end{array}$}
\end{equation}
while their ``gauge'' counterparts have the algebra
\begin{equation}\label{ward-algebra-true-new}
\fbox{$ \begin{array}{lcl}\displaystyle [\widetilde{\widetilde{{\cal W}^\Gr_{~\Dr}}} (w_1), \widetilde{\widetilde{{\cal W}^\Gr_{~\Dr}}} (w_2)] &=& \widetilde{\widetilde{{\cal W}^\Gr_{~\Dr}}} ([w_1, w_2]) + {\cal W}^\Gr_{~\Lr} (w_1 w_2 \cdot \bar{F}) \\
\displaystyle [{\cal W}^\Gr_{~\Lr} (\lambda_1), {\cal W}^\Gr_{~\Lr} (\lambda_2)] &=& {\cal W}^\Gr_{~\Lr} ([\lambda_1, \lambda_2]) \\
\displaystyle [\widetilde{\widetilde{{\cal W}^\Gr_{~\Dr}}} (w), {\cal W}^\Gr_{~\Lr} (\lambda)] &=& {\cal W}^\Gr_{~\Lr} (w \cdot \bar{\nabla} \lambda) \end{array}$}
\end{equation}
Both algebras, \eqref{ward-algebra-back-new} and \eqref{ward-algebra-true-new}, respectively, are going to become important in a moment: The ``background'' transformations and their commutators will determine the theory space on which the RG flow is taking place, while the algebra of the ``gauge'' transformations determines the ghost action \cite{ym}.

Concretely, we choose the gauge conditions to be linear in $\varepsilon^a_{~\mu}$ and independent of $\tau^{ab}_{~~\mu}$ \cite{roberto-gf}: 
\begin{subequations}\label{gc}
\begin{equation}
{\cal F}_\mu = \bar{e}_a^{~\nu} \big[\bar{D}_\nu \varepsilon^a_{~\mu} + \beta_{\rm D} \bar{D}_\mu \varepsilon^a_{~\nu}\big]\:,\label{gc-diff}
\end{equation}
\begin{equation}
{\cal G}^{ab} = \frac{1}{2}\bar{g}^{\mu\nu}\big[\varepsilon^a_{~\mu} \bar{e}^b_{~\nu} - \varepsilon^b_{~\nu} \bar{e}^a_{~\nu}\big] \equiv \varepsilon^{[ab]}\label{gc-o4}
\end{equation}
\end{subequations}
Thus, in total, there are three gauge fixing parameters: $\alpha_{\rm D}$, $\alpha_{\rm L}$ and $\beta_{\rm D}$\footnote{As can be inferred from \eqref{gf} and \eqref{gc}, the diffeomorphism gauge parameter $\alpha_{\rm D}$ is dimensionless whereas the Lorentz-gauge parameter $\alpha_{\rm L}$ is of mass dimension $-4$. Therefore, it has to be rescaled properly. We perform this rescaling by means of the mass parameter $\bar{\mu}$ that will be introduced in a moment. Within the (propertime) approximation used, no scale derivatives of dimensionless couplings appear on the right-hand side of the flow equation. Therefore, defining an additional factor of $g$ into $\alpha_{\rm L}$ will not lead to additional contributions; see the captions of Tab.\,\ref{eomega1-proceedings-1-tab} and Fig.\,\ref{eomega1-proceedings-1-tab}.}. Using \eqref{gc-diff}, \eqref{gc-o4} in \eqref{gf} we can verify that $S_{\rm gf} [\varepsilon, \tau; \bar{e}, \bar{\omega}]$ is indeed background gauge invariant:
\begin{equation}
{\cal W}^\Br_{~\Lr} \, S_{\rm gf} = 0 = \widetilde{\widetilde{{\cal W}^\Br_{~\Dr}}} \, S_{\rm gf} ~~ \Leftrightarrow ~~ {\cal W}^\Br_{~\Lr} \, S_{\rm gf} = 0 = {\cal W}^\Br_{~\Dr} \, S_{\rm gf}\:.
\end{equation}

The ghost sector requires some care, and this is indeed the reason for considering the modified diffeomorphisms. We would like the ghost action $S_{\rm gh} [\varepsilon, \tau, {\cal C}, \bar{{\cal C}}, \Sigma, \bar{\Sigma}; \bar{e}, \bar{\omega}]$ to be background gauge invariant, too. However, straightforwardly applying the Faddeev-Popov procedure to the original transformations
\begin{align}
\delta^\Gr = \left( \begin{array}{c} \delta^\Gr_{\:\Dr} (w) \\ \delta^\Gr_{\:\,\Lr} (\lambda) \end{array} \right)
\end{align} 
we obtain, in the $\bar{\Sigma}-{\cal C}$-sector, the ghost action\footnote{We consider the case $\varepsilon^a_{~\mu} = 0$ here \cite{eomega2}.} 
\begin{align}
S_{\rm gf}^{\bar{\Sigma}-{\cal C}} [{\cal C}, \bar{\Sigma}; \bar{e}, \bar{\omega}] = - \int\dr^4x\,\bar{e}\left.\left(\bar{\Sigma}_{ab} \frac{\partial {\cal G}^{ab}}{\partial \varepsilon^c_{~\nu}} \delta^\Gr_{\:\Dr} ({\cal C}) \varepsilon^c_{~\nu}\right)\right|_{\varepsilon = 0} 
\end{align}
which, with \eqref{gc-o4}, evaluates to
\begin{align}\label{sgf}
S_{\rm gf}^{\bar{\Sigma}-{\cal C}} [{\cal C}, \bar{\Sigma}; \bar{e}, \bar{\omega}] = - \int\dr^4x\,\bar{e}\:\bar{\Sigma}_{ab} \:\bar{e}^{b\mu} {\cal L}_{\cal C} \bar{e}^a_{~\mu}\:. 
\end{align}
While this functional is invariant under background diffeomorphisms, it fails to be invariant under the ${\sf O (4)}_{\rm loc}$ transformations $\delta_{\:\,\Lr}^\Br (\lambda)$, the reason being that the Lie derivative of an ${\sf O (4)}$ tensor does not define an ${\sf O (4)}$ tensor. Rather, we have ${\cal L}_{\cal C} (\lambda^a_{~b} \, \bar{e}^b_{~\mu}) \neq \lambda^a_{~b} \, {\cal L}_{\cal C} \bar{e}^b_{~\mu}$, since $\lambda^a_{~b} (x)$ is a spacetime scalar which transforms nontrivially under diffeomorphisms. Stated differently, ${\sf O (4)}_{\rm loc}$ transformations and (ordinary) diffeomorphisms do not commute, and this is exactly what the above algebra relations express.

The way out consists in applying the Faddeev-Popov procedure to the ${\sf O (4)}_{\sf loc}$-covariantized (true) gauge transfomations 
\begin{align}
\widetilde{\widetilde{\delta^\Gr}} = \left( \begin{array}{c} \widetilde{\widetilde{\delta^\Gr_{\:\Dr}}} (w) \\ \delta_{\:\,\Lr}^\Gr (\lambda) \end{array} \right) \:.
\end{align}  
They are broken by the ten gauge fixing conditions 
\begin{align}
\left( \begin{array}{c} {\cal F}_\mu \\ {\cal G}^{ab} \end{array} \right) \equiv \big( {\cal Q}^I \big)
\end{align}
for which we use a uniform notation where $\big({\cal Q}^I\big) \equiv \big({\cal F}_\mu\big)$ for $I = 1, \cdots, 4$ and $\big({\cal Q}^I\big) \equiv \big({\cal G}^{ab}\big)$ for $I = 5, \cdots, 10$. Denoting, in the same fashion, the ten parameters of the gauge transformations as $\big( \Lambda^I \big)  = \big(w^\mu, \lambda^a_{~b}\big)$, the Faddeev-Popov determinant reads 
\begin{align}
{\rm det} \left. \left(\frac{\delta {\cal Q}^I (x)}{\delta \Lambda^J (y)}\right)\right|_{\Lambda = 0}\:,
\end{align}
and exponentiating it we obtain a ghost action which has the structure
\begin{align}
- \int\dr^4 x \:\bar{e} \left( \begin{array}{c} \bar{{\cal C}_\mu} \\ \bar{\Sigma}_{ab} \end{array} \right)^{\rm T} \left(\begin{array}{cc} \Omega^\mu_{~\nu} & \Omega^\mu_{~cd} \\ \Omega^{ab}_{~~\nu} & \Omega^{ab}_{~~cd} \end{array} \right) \left( \begin{array}{c} {\cal C}^\nu \\ \Sigma^{cd} \end{array} \right) 
\end{align}
The Faddeev-Popov operator $\Omega$ is rather complicated; here we must refer to \cite{eomega2} for its explicit form. Suffice it to say that one can now check that he ghost action obtained is indeed invariant under background gauge transformations:
\begin{equation}
{\cal W}^\Br_{~\Lr} \, S_{\rm gh} = 0 = \widetilde{\widetilde{{\cal W}^\Br_{~\Dr}}} \, S_{\rm gh} ~~ \Leftrightarrow ~~ {\cal W}^\Br_{~\Lr} \, S_{\rm gh} = 0 = {\cal W}^\Br_{~\Dr} \, S_{\rm gh}\:.
\end{equation}
This property is the main prerequisite for arriving at a background gauge invariant effective average action.

\subsection{Theory space and flow equation}\label{ss23}

The functional integral \eqref{z-functional} gives rise to the associated effective average action \cite{mr} in the standard way: one adds a $\delta^{\rm B}$-invariant mode cutoff to the bare action, 
\begin{align}
\Delta_k S \propto \int {\rm d}^4 x\:\bar{e}\:(\varepsilon, \tau)\,{\cal R}_k\,(\varepsilon, \tau)^{\rm T}\,,
\end{align} 
\noindent couples $\varepsilon$ and $\tau$ to sources, Legendre transforms the resulting generating functional ${\rm ln}\,{\cal Z}$, and finally subtracts $\Delta_k S$ for the expectation value field in order to arrive at the running action: 
\begin{align}
\Gamma_k [\varepsilon, \tau, \xi, \bar{\xi}, \Upsilon, \bar{\Upsilon}; \bar{e}, \bar{\omega}] \equiv \Gamma_k [e, \omega, \bar{e}, \bar{\omega}, \xi, \bar{\xi}, \Upsilon, \bar{\Upsilon}]\,. 
\end{align}
\noindent Therein $\bar{\varepsilon}^a_{~\mu}$, $\bar{\tau}^{ab}_{~~\mu}$ as well as 
\begin{align}
e^a_{~\mu} \equiv \langle \hat{e}^a_{~\mu} \rangle = \bar{e}^a_{~\mu} + \bar{\varepsilon}^a_{~\mu}
\end{align} 
and 
\begin{align}
\omega^{ab}_{~~\mu} \equiv \langle \hat{\omega}^{ab}_{~~\mu} \rangle = \bar{\omega}^{ab}_{~~\mu} + \bar{\tau}^{ab}_{~~\mu}
\end{align} 
denote the expectation value fields, while we write for the ghosts
\begin{align}
\xi^\mu \equiv \langle {\cal C}^\mu \rangle \:,~ \bar{\xi}_\mu \equiv \langle \bar{{\cal C}}_\mu \rangle \:,~ \Upsilon^{ab} \equiv \langle \Sigma^{ab} \rangle \:,~\bar{\Upsilon}_{ab} \equiv \langle \bar{\Sigma}_{ab} \rangle \:.
\end{align}
The average action $\Gamma_k$ may be considered a functional of either the fluctuations $\bar{\varepsilon}^a_{~\mu}$ and $\bar{\tau}^{ab}_{~~\mu}$ or the complete classical fields $e^a_{~\mu}$ and $\omega^{ab}_{~~\mu}$. 

Obviously the action $\Gamma_k$ is defined on a rather complicated theory space ${\cal T}$ consisting of functionals depending on two independent vielbein variables $(e, \,\bar{e})$, two spin connections $(\omega, \,\bar{\omega})$, as well as on the diffeomorphism and ${\sf O}(4)$ ghosts and antighosts, respectively. The functionals in ${\cal T}$ are constrained by the requirement of background gauge invariance:
\begin{align}
{\cal T} \equiv \Big\{F\,\big|\,{\cal W}^{\rm B}_{~\Dr} (w)\,F = 0 ~ \wedge ~ {\cal W}^{\rm B}_{~\Lr} (\lambda) \,F = 0 ~~ \forall ~~ w^\mu,\:\lambda^{ab}\Big\}\:.
\end{align}

From the above functional integral based construction of $\Gamma_k$ one straightforwardly derives the FRGE it satisfies: 
\begin{align}\partial_k \Gamma_k = \frac{1}{2} {\rm STr} \big[(\Gamma_k^{(2)} + {\cal R}_k)^{-1} \partial_k {\cal R}_k\big]\:.
\end{align} 
With the kernel ${\cal R}_k [\bar{e}, \bar{\omega}]$ specified appropriately, the equation indeed defines a flow on ${\cal T}$, i.\,e. it does not generate background gauge invariance violating terms.

%
%
%
\section{Results}\label{s3}
We have solved the flow equation for $\Gamma_k [e, \omega, \cdots]$ on a three-dimensional truncated theory space spanned by actions of the Holst type:
\begin{eqnarray}
\Gamma_k [e, \omega, \cdots] &=& - \frac{1}{16 \pi G_k} \int {\rm d}^4 x\:e\,\Big[e_a^{~\mu} e_b^{~\nu}\Big(F^{ab}_{~~\mu\nu} - \frac{1}{\gamma_k} \star F^{ab}_{~~\mu\nu}\Big)\nonumber\\
&& \hspace{2.9cm} - 2\:\Lambda_k\Big] + S_{\rm gf} + S_{\rm gh} \label{holst-truncation}
\end{eqnarray}
In practice we used, because of the enormous algebraic complexity of the calculations involved, a slightly simplified version of the FRGE of the propertime type. An equation of the same type has been used within the Einstein-Hilbert truncation of metric gravity \cite{prop}, and virtually the same results were found as with the exact RG equation in this truncation. 

The truncation ansatz \eqref{holst-truncation} consists of the Hilbert-Palatini action known from Einstein-Cartan gravity, plus the Immirzi term which only exists in four dimensions; in fact $\star F^{ab}_{~~\mu\nu} \equiv \frac{1}{2} \varepsilon^{ab}_{~~cd} F^{cd}_{~~\mu\nu}$ is the dual of the curvature of $\omega$, $F \equiv F (\omega)$, with respect to the frame indices. Besides $G_k$, \eqref{holst-truncation} contains two more running parameters: the cosmological constant $\Lambda_k$ and the Immirzi parameter $\gamma_k$. The gauge fixing and ghost terms are assumed to retain their classical form for all $k$, except for the replacement $G \to G_k$. The parameters $\alpha_{\rm D}$, $\alpha_{\rm L}$ and $\beta_{\rm D}$ are treated as constant in the approximation considered. Thus the truncated theory space can be coordinatized by a triple $(g, \lambda, \gamma)$ where $g_k \equiv G_k\,k^2$ and $\lambda_k \equiv \Lambda_k / k^2$ are the dimensionless Newton's and cosmological constant, respectively.  

With $t \equiv {\rm ln}\,k$, the RG equations are of the form $\partial_t g_k = \beta_g \equiv (2 + \eta_{\rm N})g_k, \:\partial_t \lambda_k = \beta_\lambda, \:\partial_t \gamma_k = \beta_\gamma$ where the anomalous dimension of Newton's constant, $\eta_{\rm N}$, and the other beta functions are given by
\begin{eqnarray}
\eta_{\rm N} (g, \lambda, \gamma) &=& 16 \pi \, g \, f_+ (\lambda, \gamma)\nonumber\\
\beta_\gamma (g, \lambda, \gamma) &=& 16 \pi \, g \, \gamma \Big[\gamma\,f_- (\lambda, \gamma) -f_+ (\lambda, \gamma)\Big]\label{flow-exact1}\\
\beta_\lambda (g, \lambda, \gamma) &=& -2\,\lambda + 8 \pi \, g \Big[2\,\lambda \, f_+ (\lambda, \gamma) + f_3 (\lambda, \gamma)\Big]\nonumber
\end{eqnarray}
The functions $f_\pm$ and $f_3$ are extremely complicated and cannot be written down here. Besides $\lambda$ and $\gamma$, they depend parametrically also on the three gauge fixing parameters and an additional parameter $\bar{\mu}$ with the dimension of a mass. The latter is needed to give a uniform dimension to the fluctuations $\bar{\varepsilon}^a_{~\mu}$ and $\bar{\tau}^{ab}_{~~\mu}$: Only after a rescaling of the form $\bar{\varepsilon}^a_{~\mu} \rightarrow \bar{\mu}^{\frac{1}{2}}\,\bar{\varepsilon}^a_{~\mu}$, $\bar{\tau}^{ab}_{~~\mu} \rightarrow \bar{\mu}^{ - \frac{1}{2}}\,\bar{\tau}^{ab}_{~~\mu}$ the effective inverse propagator $\Gamma_k^{(2)}$ constitutes an operator with a welldefined spectrum and a welldefined trace. The parameter $\bar{\mu}$ might be treated as a running quantity in a more complete truncation.     

We coordinatize ${\cal T}$ by an atlas consisting of two charts. In order to cover the neighborhood of the submanifold $\gamma = \pm \infty$ in ${\cal T}$, we introduce a new coordinate $\hat{\gamma}$. In the overlap $|\gamma| \in ~ ]0, +\infty[$ of the $(g, \lambda, \gamma)\,$- and the $(g, \lambda, \hat{\gamma})\,$-chart, the coordinates $\gamma$ and $\hat{\gamma}$ are related by the transition function $\hat{\gamma} (\gamma) = \gamma^{-1}$. With $\beta_{\hat{\gamma}} (g, \lambda, \hat{\gamma}) = - {\hat{\gamma}}^2\,\beta_\gamma (g, \lambda, {\hat{\gamma}}^{-1})$, the flow equation in the $\hat{\gamma}$-chart is given by
\begin{eqnarray}
\eta_{\rm N} (g, \lambda, \gamma) &=& 16 \pi \, g \, f_+ (\lambda, {\hat{\gamma}}^{-1})\nonumber\\
\beta_{\hat{\gamma}} (g, \lambda, \hat{\gamma}) &=& 16 \pi \, g \, \hat{\gamma} \Big[f_+ (\lambda, {\hat{\gamma}}^{-1}) - {\hat{\gamma}}^{-1}\,f_- (\lambda, {\hat{\gamma}}^{-1})\Big]\label{flow-exact2}\\
\beta_\lambda (g, \lambda, \hat{\gamma}) &=& - 2\,\lambda + 8 \pi \, g \Big[2\,\lambda \, f_+ (\lambda, {\hat{\gamma}}^{-1}) + f_3 (\lambda, {\hat{\gamma}}^{-1})\Big]\nonumber
\end{eqnarray}

We studied the system \eqref{flow-exact1}, \eqref{flow-exact2} for various propertime cutoff functions and gauge parameters $\alpha_{\rm D}$, $\alpha_{\rm L}$ and $\beta_{\rm D}$. Within our approximation, $\mu \equiv \bar{\mu} / k$ can be assumed to be a constant, $k$-independent number. 

For $\mu \gtrsim 2$ the RG flow displays the following features: 

\noindent {\bf (i)} It is reflection symmetric under $\gamma \to - \gamma$. 

\noindent {\bf (ii)} The beta-functions $\beta_g$, $\beta_\gamma$, $\beta_{\hat{\gamma}}$ and $\beta_\lambda$ contain simple poles at $\gamma = \hat{\gamma} = \pm 1$. However, those  are presumably an artifact of the approximation of the exact flow that we employed. In fact, our analysis strongly suggests that for $\gamma$ not too close to $\pm 1$ the functions $f_\pm$ and $f_3$ are actually {\em independent} of $\gamma$. For such values of $\gamma$ it is a rather precise approximation to replace them by functions $\tilde{f}_\pm$ and $\tilde{f}_3$ that only depend on $\lambda$, leading to the simpler system
\begin{eqnarray}
\partial_t\,g_k &=& \Big[2 + 16 \pi \, g_k \, \tilde{f}_+ (\lambda_k)\Big] g_k\nonumber\\
\partial_t\,\gamma_k &=& 16 \pi \, g_k \, \gamma_k \Big[\gamma_k\,\tilde{f}_- (\lambda_k) - \tilde{f}_+ (\lambda_k)\Big]\label{flow-hypo1}\\
\partial_t\,\lambda_k &=& - 2\,\lambda_k + 8 \pi \, g_k \Big[2\,\lambda_k \, \tilde{f}_+ (\lambda_k) + \tilde{f}_3 (\lambda_k)\Big]\nonumber
\end{eqnarray}      
and likewise for the $\hat{\gamma}$-chart. While the equations \eqref{flow-hypo1} are equivalent to \eqref{flow-exact1} when $|\gamma| \not\approx 1$, a detailed analysis \cite{eomega2} indicates that for $|\gamma| \to 1$, too, the regular beta functions \eqref{flow-hypo1} rather than those of \eqref{flow-exact1} are likely to apply.

\noindent {\bf (iii)} The system \eqref{flow-hypo1} and its analogue in the $\hat{\gamma}$-chart imply $\beta_\gamma = 0$ and $\beta_{\hat{\gamma}} = 0$ for $\gamma^\star = 0$ and $\hat{\gamma}^\star = 0$, respectively. For each of the two sets of equations we find a fixed point $\mbox{{\bf NGFP}}_{\boldsymbol{0}} \equiv (g^\star_0, \lambda^\star_0, \gamma^\star)$ and $\mbox{{\bf NGFP}}_{\boldsymbol{\infty}} \equiv (g^\star_\infty, \lambda^\star_\infty, \hat{\gamma}^\star)$ of \eqref{flow-exact1}, \eqref{flow-exact2} with $g^\star_{0, \infty} > 0$, $\lambda^\star_{0, \infty} < 0$ and $g^\star_0 \neq g^\star_\infty$, $\lambda^\star_0 \neq \lambda^\star_\infty$. For the choice $\mu = 5$ we obtained the values shown in Tab.\,\ref{eomega1-proceedings-1-tab}.
\begin{table}
\begin{center}
\begin{tabular}{|c|c|c|c|c|c|c|}\hline
$\mbox{{\bf NGFP}}_{\boldsymbol{0}}$ & $g^\star_0$ & $\lambda^\star_0$ & $g^\star_0\,\lambda^\star_0$ & $\Theta_1$ &$\Theta_2$ & $\Theta_\gamma$ \\ \hline \hline
$\alpha_{\rm D} = 1$ & 3.37 & -6.78 & -22.86 & 1.94 & 3.71 & -1.98 \\ \hline
$\alpha_{\rm D} = 10$ & 1.36 & -1.08 & -1.47 & 2.46 & -6.64 & -0.43 \\ \hline
$\alpha_{\rm D} = 0.1$ & 3.65 & -7.42 & -27.09 & 2.28 & 3.73 & -2.00 \\ \hline\hline
$\mbox{{\bf NGFP}}_{\boldsymbol{\infty}}$ & $g^\star_\infty$ & $\lambda^\star_\infty$ & $g^\star_\infty\,\lambda^\star_\infty$ & $\Theta_1$ &$\Theta_2$ & $\Theta_{\hat{\gamma}}$ \\ \hline \hline
$\alpha_{\rm D} = 1$ & 3.30 & -4.18 & -13.79 & 1.81 & 3.22 & 1.94 \\ \hline
$\alpha_{\rm D} = 10$ & 2.18 & -1.83 & -3.98 & 2.76 & -2.40 & 1.34 \\ \hline
$\alpha_{\rm D} = 0.1$ & 3.86 & -5.16 & -19.89 & 2.55 & 3.32 & 2.01 \\ \hline
\end{tabular}
\caption{Properties of the fixed points $\mbox{{\bf NGFP}}_{\boldsymbol{0}}$ and $\mbox{{\bf NGFP}}_{\boldsymbol{\infty}}$ of the $(g, \lambda, \gamma)\,$- and the $(g, \lambda, \hat{\gamma})\,$-system, respectively. The numerical values were obtained for the choice $\mu = 5$, $\beta_{\rm D} = 0$, $\alpha_{\rm L} = 16 \pi g\,{\bar{\mu}}^{-4}$ with a sharp propertime cutoff.}
\label{eomega1-proceedings-1-tab}
\end{center}
\end{table} 

\noindent {\bf (iv)} At both fixed points, the $g$ and $\lambda$ directions are to a very good approximation eigendirections of the linearized flow on ${\cal T}$, whereas this is exactly true for the $\gamma$- and $\hat{\gamma}$-directions, respectively. At $\mbox{{\bf NGFP}}_{\boldsymbol{0}}$ and $\mbox{{\bf NGFP}}_{\boldsymbol{\infty}}$, both the $g$ and $\lambda$ directions are relevant scaling fields, i.\,e. their associated critical exponents $\Theta_1$ and $\Theta_2$ are real and positive. In contrast, at $\mbox{{\bf NGFP}}_{\boldsymbol{0}}$ the Immirzi parameter $\gamma$ is irrelevant ($\Theta_\gamma < 0$), whereas at $\mbox{{\bf NGFP}}_{\boldsymbol{\infty}}$ its inverse $\hat{\gamma}$ is relevant ($\Theta_{\hat{\gamma}} > 0$).  

\begin{figure*}[htp]
\psfrag{g}[bl][bl]{$g$}
\psfrag{l}[bl][bl]{$\lambda$}
\psfrag{G}[bl][bl]{$\gamma$}
\psfrag{Gh}[bl][bl]{$\hat{\gamma}$}
\begin{center}
\subfigure[The $(g, \lambda)\,$-section of the flow at $\mbox{{\bf NGFP}}_{\boldsymbol{0}}$.]{\label{eomega1-proceedings-1a-fig}\includegraphics[scale=0.75]{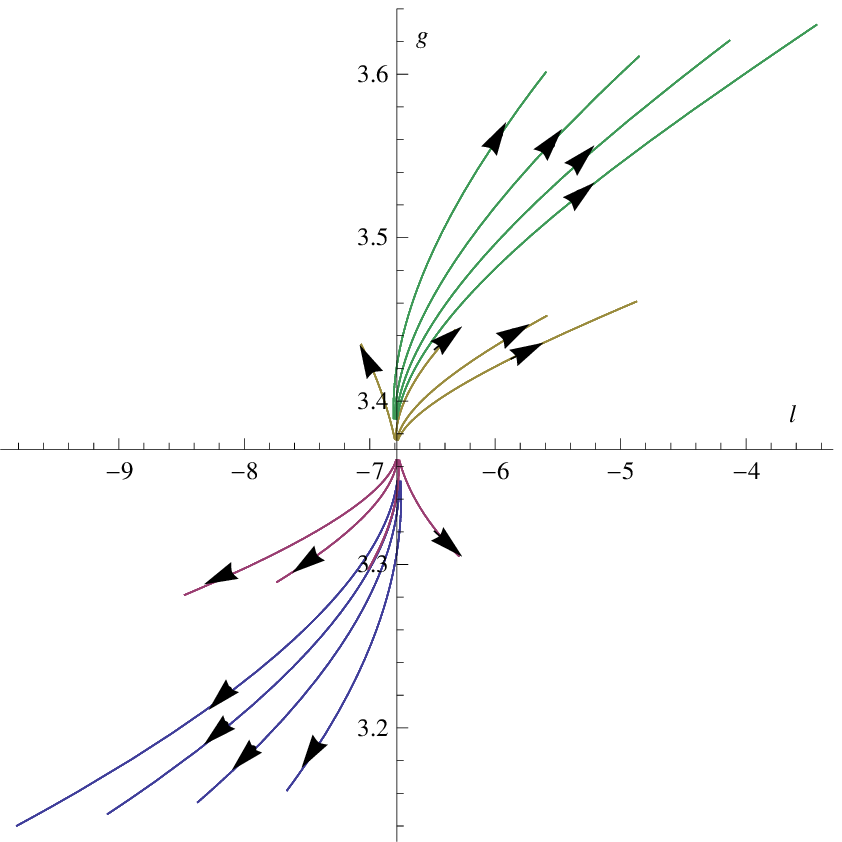}}
\subfigure[The $(g, \lambda)\,$-section of the flow at $\mbox{{\bf NGFP}}_{\boldsymbol{\infty}}$.]{\label{eomega1-proceedings-1b-fig}\includegraphics[scale=0.75]{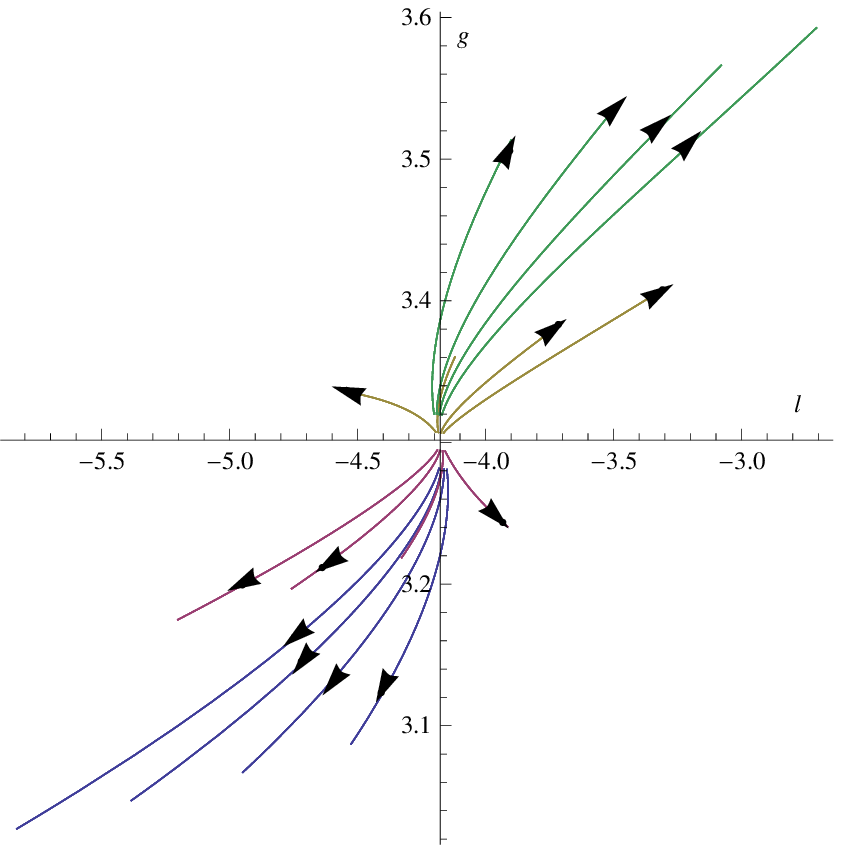}}
\subfigure[The $(g, \gamma)\,$-section of the flow at $\mbox{{\bf NGFP}}_{\boldsymbol{0}}$.]{\label{eomega1-1c-fig}\includegraphics[scale=0.75]{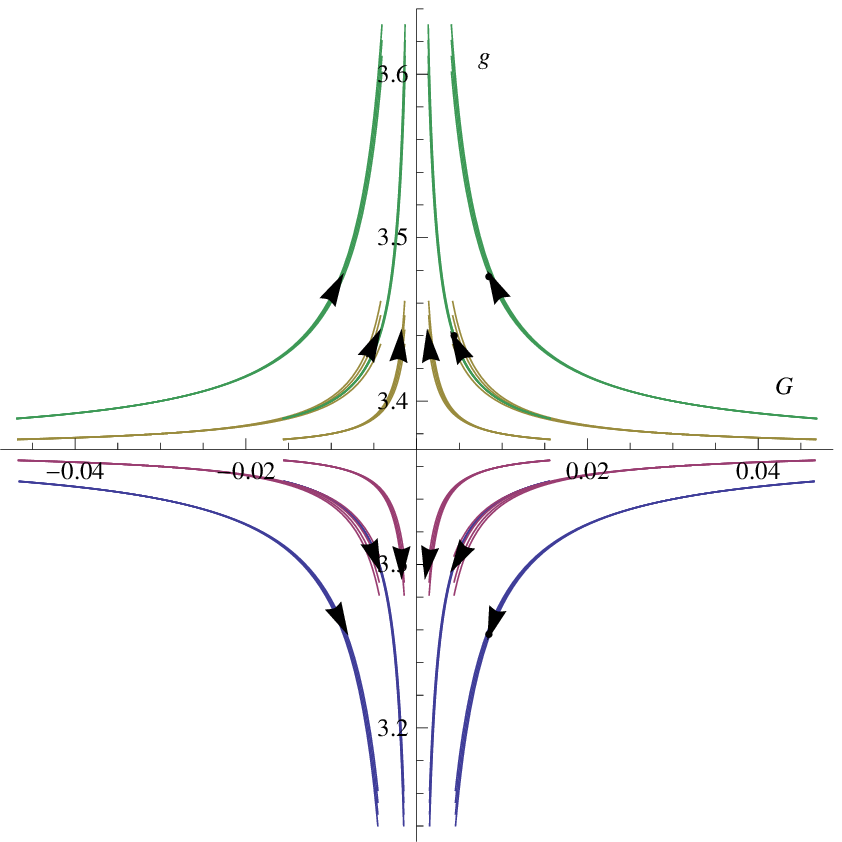}}
\subfigure[The $(g, \hat{\gamma})\,$-section of the flow at $\mbox{{\bf NGFP}}_{\boldsymbol{\infty}}$.]{\label{eomega1-1d-fig}\includegraphics[scale=0.75]{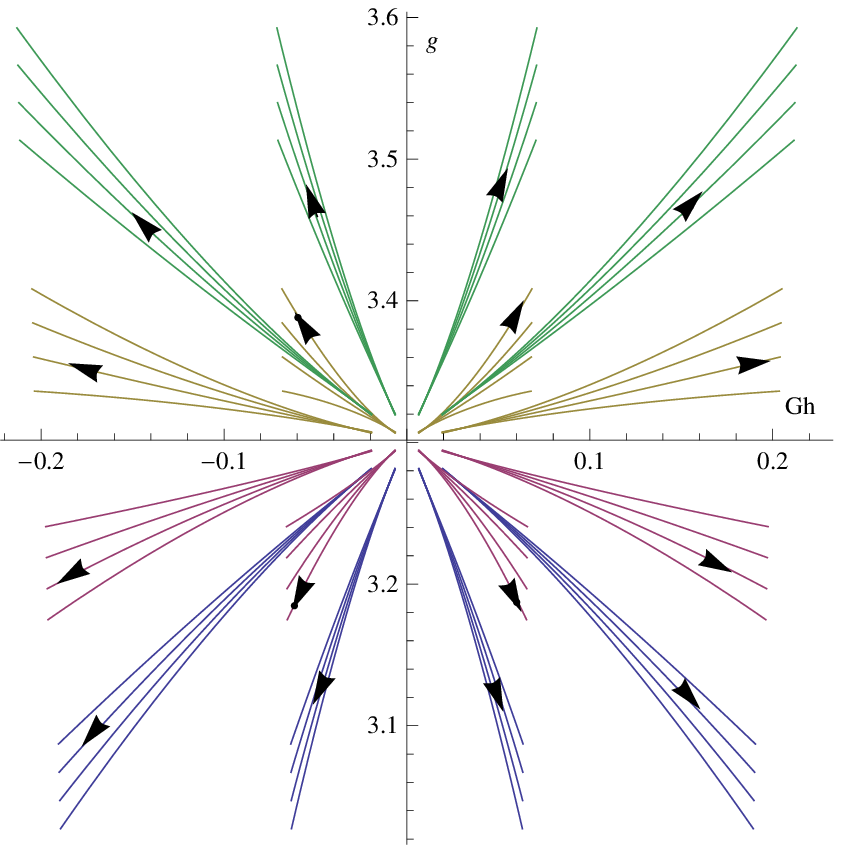}}
\subfigure[The $(\gamma, \lambda)\,$-section of the flow at $\mbox{{\bf NGFP}}_{\boldsymbol{0}}$.]{\label{eomega1-proceedings-1e-fig}\includegraphics[scale=0.75]{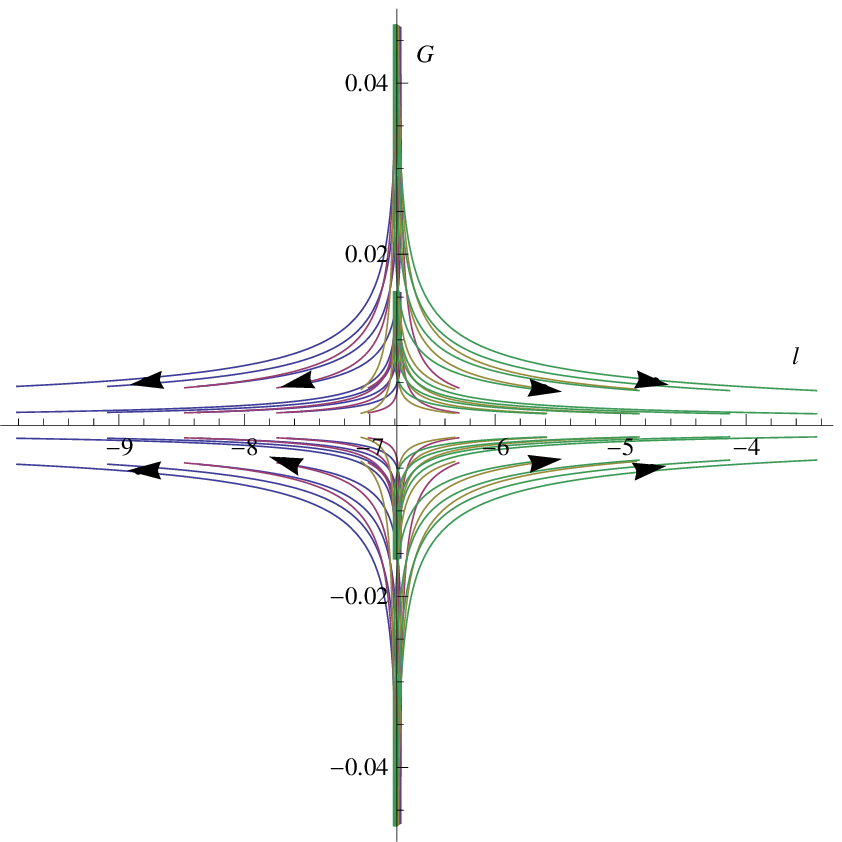}}
\subfigure[The $(\hat{\gamma}, \lambda)\,$-section of the flow at $\mbox{{\bf NGFP}}_{\boldsymbol{\infty}}$.]{\label{eomega1-proceedings-1f-fig}\includegraphics[scale=0.75]{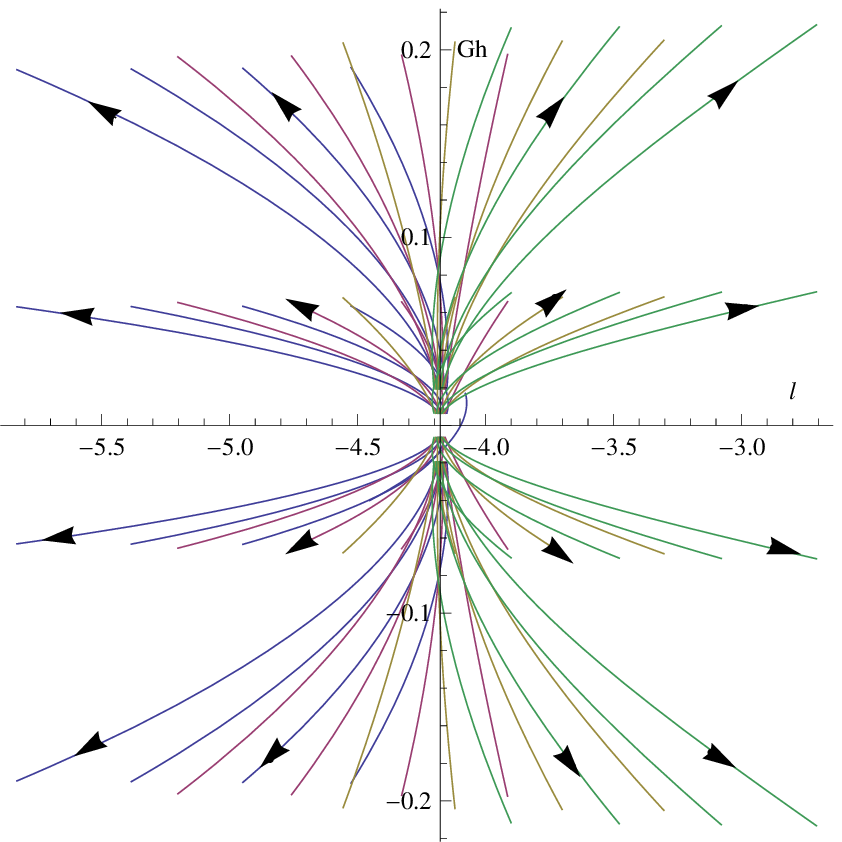}}
\end{center}
\caption{The two-dimensional sections of the flow near $\mbox{{\bf NGFP}}_{\boldsymbol{0}}$ and $\mbox{{\bf NGFP}}_{\boldsymbol{\infty}}$, respectively, for the choice $\mu = 5$, $\alpha_{\rm D} = 1$, $\beta_{\rm D} = 0$, $\alpha_{\rm L} = 16 \pi g\,{\bar{\mu}}^{-4}$ with a sharp propertime cutoff. The arrows point in the direction of decreasing $k$.}
\label{eomega1-proceedings-1-fig}
\end{figure*}

\noindent {\bf (v)} The three two-dimensional sections of the flow at each fixed point  are presented in Fig.\,\ref{eomega1-proceedings-1-fig}. These plots were obtained for the choice $\mu = 5$, $\alpha_{\rm D} = 1$, $\beta_{\rm D} = 0$, $\alpha_{\rm L} = 16 \pi g\,{\bar{\mu}}^{-4}$ and by means of a sharp propertime cutoff.  

\noindent {\bf (vi)} By letting $\gamma \to \infty$ in \eqref{holst-truncation} we can study the two-dimensional $(g, \lambda)\,$-truncation, i.\,e. the Hilbert-Palatini truncation, which is {\em not} equivalent to the Einstein-Hilbert truncation of QEG. In this case, we find a fixed point $(g^\star, \lambda^\star)$ with $g^\star > 0$, $\lambda^\star < 0$ that exhibits the same features as the $(g, \lambda)\,$-sections of the two fixed points $\mbox{{\bf NGFP}}_{\boldsymbol{0}}$ and $\mbox{{\bf NGFP}}_{\boldsymbol{\infty}}$ of the three-dimensional truncation. In particular, the flow basically looks like the ones depicted in Fig.\,\ref{eomega1-proceedings-1a-fig} and Fig.\,\ref{eomega1-proceedings-1b-fig}, respectively. 

\noindent {\bf (vii)} For $\lambda = 0$, we obtain the $(g, \gamma)\,$- and $(g, \hat{\gamma})\,$-truncation, respectively, with $\beta_g$, $\beta_\gamma$ and $\beta_{\hat{\gamma}}$ given by \eqref{flow-exact1} and \eqref{flow-exact2}, but with the functions $f_\pm$ evaluated at $\lambda = 0$. In this case our results are compatible with $\beta_\gamma = 0 = \beta_{\hat{\gamma}} \Leftrightarrow f_+ (0, \gamma)|_{\gamma = 0,\,\pm\infty} = \big(\gamma\,f_- (0, \gamma)\big)|_{\gamma = 0,\,\pm\infty}$, i.\,e. the renormalization of the remaining invariant would solely be given by the renormalization of Newton's constant. While this result needs to be corroborated by a more pecise treatment, we find $g^\star|_{\gamma^\star = 0} = g^\star|_{\hat{\gamma}^\star = 0} > 0$ within this truncation. These investigations suggest that the Immirzi parameter owes its RG running to a nonzero cosmological constant. 

\noindent {\bf (viii)} With respect to variations of the regularization scheme our results are remarkably robust. The signs of the fixed point coordinates and of most of the quantities that are expected to be universal are gauge parameter independent, as well. Nevertheless, the quantitative gauge parameter dependence of the universal quantities such as the product $g^\star_{0, \infty} \, \lambda^\star_{0, \infty}$ and the critical exponents is somewhat stronger than in comparable calculations within metric gravity \cite{prop}.

%
%
%
\section{Conclusion}\label{s4}
We find significant evidence for Asymptotic Safety of pure gravity in the Einstein-Cartan approach. There seem to exist two NGFPs, located at $\gamma = 0$ and $\gamma = \pm \infty$, which in principle both are suitable for taking the continuum limit there. By investigating how observables depend upon $\gamma$ in particular, one may determine the physical properties of the resulting quantum field theories. Using either fixed point for the Asymptotic Safety construction, gravity is anti-screening in the UV, i.\,e. $g^\star_{0, \infty} > 0$, but in contrast to QEG the cosmological constant is negative in the fixed point regime, $\lambda^\star_{0, \infty} < 0$. However, this does not contradict present day observations since $\lambda$ might very well flow to positive values for IR scales of the order of typical astronomical distances. Future investigations should aim at a better control of the gauge dependencies and at understanding the phenomenological implications of the scale dependent Immirzi parameter.

%
%
%
%

%
%

%

%
%
%
%
%
%
%


\begin{thebibliography}{99}
%
\bibitem{mr}
M.~Reuter, Phys.\ Rev.\ D 57 (1998) 971 and hep-th/9605030.
%
\bibitem{wein}
S.~Weinberg 
in {\it General Relativity, an Einstein Centenary Survey,}
S.W.~Hawking and W.~Israel (Eds.), Cambridge University Press, Cambridge (1979).
%
\bibitem{QEG}
O.~Lauscher and M.~Reuter, 
Phys.\ Rev.\ D 65 (2002) 025013 and hep-th/0108040;\\
M.~Reuter and F.~Saueressig, 
Phys.\ Rev.\ D 65 (2002) 065016 and hep-th/0110054;\\
O.~Lauscher and M.~Reuter, 
Phys.\ Rev.\ D 66 (2002) 025026 and hep-th/0205062;\\
O.~Lauscher and M.~Reuter, 
Class.\ Quant.\ Grav.\ 19 (2002) 483 and hep-th/0110021;\\
A.~Codello, R.~Percacci, and C.~Rahmede,
Ann.\ Phys.\ 324 (2009) 414;\\
D.~Benedetti, P.~Machado, and F.~Saueressig, 
Nucl.\ Phys.\ B824 (2010) 168;\\
D.~Benedetti, K.~Groh, P.~Machado, F.~Saueressig, arXiv:1012.3081 [hep-th].
%
\bibitem{prop}
A.~Bonanno and M.~Reuter,
JHEP 02 (2005)035 and hep-th/0410191.
%
\bibitem{reviews}
For reviews see: 
M.~Reuter and F.~Saueressig 
in {\it Geometric and Topological Methods for Quantum Field Theory,} H.~Ocampo, S.~Paycha, and A.~Vargas (Eds.), Cambridge University Press, Cambridge (2010) and arXiv:0708.1317 [hep-th];\\
M.~Niedermaier and M.~Reuter,
Living Rev.\ in Relativity 9 (2006) 5;\\
R.~Percacci 
in {\it Approaches to Quantum Gravity,} D.~Oriti (Ed.), Cambridge University Press, Cambridge (2009).
%
\bibitem{elisa}
E.~Manrique and M.~Reuter, Phys.\ Rev.\ D 79 (2009) 025008 and arXiv:0811.3888 [hep-th].
%
\bibitem{A}
A.~Ashtekar, 
{\it Lectures on non-perturbative canonical gravity,} World Scientific, Singapore (1991);\\ 
A.~Ashtekar and J.~Lewandowski, 
Class.\ Quant.\ Grav. 21 (2004) R53.
%
\bibitem{R}
C.~Rovelli, 
{\it Quantum Gravity,} Cambridge University Press, Cambridge (2004).
%
\bibitem{T}
Th.~Thiemann, 
{\it Modern Canonical Quantum General Relativity,} Cambridge University Press, Cambridge (2007).
%
\bibitem{SF}
A.~Perez, 
Class.\ Quant.\ Grav. 20 (2003) R43.
%
\bibitem{GFT}
D.~Oriti
in {\it Approaches to Quantum Gravity,} D.~Oriti (Ed.), Cambridge University Press, Cambridge (2009);\\ 
L.~Freidel, 
Int.\ J.\ Theor.\ Phys. 44 (2005) 1769.
%
\bibitem{soeren}
S.~Holst, 
Phys.\ Rev.\ D 53 (1996) 5966.
%
\bibitem{eomega2}
J.-E.~Daum and M.~Reuter, 
preprint arXiv:1012.4280 [hep-th] and in preparation.
%
%
%
%
%
\bibitem{ym}
J.-E.~Daum, U.~Harst, and M.~Reuter, 
JHEP 01 (2010) 084 and arXiv:0910.4938 [hep-th]. 
%
\bibitem{roberto-gf}
R.~Floreanini and R.~Percacci 
in {\it Gravitation Theory and Geometric Methods in Field Theory, Volume in honour of D.~Ivanenko's 90th jubilee,} V.~Koloskov (Ed.), Moscow (1994).
%
%
%
%
%
%
%
%
%
%
%
%
%
%
%
%
%
%
%
%
%
%
%
%
%
%
%
%
%
%
%
%
%
%
%
%
\end{thebibliography}
\end{document}